# Computational investigation on the thermodynamics of $H_2CO + NH_2 \rightarrow NH_2CHO + H$ on interstellar water ice surfaces

Berta Martínez-Bachs, Albert Rimola

Departament de Química, Universitat Autònoma de Barcelona, 08193 Bellaterra, Catalonia, Spain

**Abstract.** Formamide has a key role in prebiotic chemistry as it is the simplest molecule containing the four most important atoms from a biological point of view: hydrogen, carbon, nitrogen and oxygen. Due to its importance, the formation of this molecule has been studied and different pathways have been considered both in gas-phase and on ices of dust grains since it was first detected. In the present work, the thermodynamics of the formation route of formamide starting from $NH_2$ and $H_2CO$, a reaction channel proposed to occur in the gas phase, has been theoretically investigated in the scenario taking place on icy dust grains modelled by both a cluster and a periodic approach. Different DFT functionals have been employed to obtain accurate energy values for the mechanistic steps involved in the reaction.

## 1    Introduction

Interstellar complex organic molecules (iCOMs) are carbon-bearing molecules containing at least six atoms that have been detected in the interstellar medium (ISM). iCOMs are a turn out in chemical complexity and diversity, being the next group of chemical compounds after the simplest inorganic species detected [1], [2].

Formamide ($NH_2CHO$) was first detected in 1971 in the ISM [3]. It is not only interesting due to its iCOM nature but also because it is the simplest molecule that contains the four most important elements from a biological point of view, hydrogen, carbon, nitrogen and oxygen, and the simplest molecule containing the amide bond O-C-NH, the group that join amino acids forming peptides. This turn out in chemical complexity increases its interest from the standpoint of prebiotic chemistry [4], [5].

Different pathways for the formation of formamide have been discussed in the recent years, including routes in the gas phase and on the surfaces of icy dust grains, which have conducted a vivid debate. The formation of formamide on ice mantles has been studied by reaction of the radical CN with $H_2O$ molecules of the ice itself [4]. Other on-grain routes postulate the radical-radical coupling between $NH_2$ and HCO [6] adopting the general scheme for iCOM formation recurrently used in gas-grain astrochemical modelling [7].

The formation of formamide has also been analysed from gas-phase scenarios, considering its formation by reaction between $NH_2$ and $H_2CO$ [5]. In this route, the pathway goes through the formation of a radical intermediate $NH_2CH_2O$ which dissociates to give formamide and hydrogen, i.e.,:

$$NH_2 + CH_2O \rightarrow NH_2CH_2O \rightarrow NH_2CHO + H$$

In this mechanism, the intermediate was found to be the most stable species, i.e., the energetics of the overall reaction are disfavoured because the products are more unstable than the intermediate. However, since the reaction is associated with the release of a H atom, kinetic calculations based on the RRKM approach indicated a relatively fast overall reaction highlighting its feasibility in the ISM. This formation route, however, has not been considered on the ice mantles. Interestingly, the interaction of the reactive species with the surfaces can modify the energetic features of the reactions, by increasing/decreasing the energy barriers and favouring/disfavouring their thermodynamics.

In this work, we address this formamide formation route occurring on interstellar water ice mantles considering only its thermodynamics, that is, accounting for the relative energies of the different minima stationary points (i.e., the reactants, the intermediate and the products). This has been studied by means of quantum chemical calculations according to different situations: in the absence of water ice and in their presence in order to determine whether water ice infers any influence on the energetics of the reaction. Water ice surfaces have been modelled using two different approaches: a cluster consisting of three water molecules, and a periodic water ice surface of a crystalline system. For all the cases, different DFT methods have been employed to check their accuracy with the goal to identify the most accurate one (with respect to CCSD(T) results), which will be used in future studies including the kinetics aspects.

## 2  Methods

Molecular calculations have been performed with the Gaussian16 software package, while periodic simulations with the CRYSTAL17 program. Both codes use gaussian functions centred to the atoms as basis set, in this way ensuring a reasonable comparison of the results.

For both molecular and periodic calculations, the formation of formamide has been analysed by optimising the structures of the reactants, intermediate and products. The following DFT functionals have been used in these geometry optimisations: the pure gradient generalized approximations (GGA) PBE, the non-local B3LYP hybrid functional that includes a 20% of exact exchange in its definition, and the meta-hybrid M06-2X functional, which incorporate a 45% of exact exchange. To take into account dispersive forces, for all the DFT functionals, the D3-based Grimme's correction term was added: for B3LYP and PBE the D3 Becke-Johnson (D3(BJ)) correction, while for M06-2X the zero-damping (D3), since the BJ parameters are not computed for this functional [8]–[13]. The basis sets used among these DFT calculations were: 6-311+G(d,p) for molecular systems, and 6-311G* for periodic systems.

Coupled Cluster (CC) is a family of the most accurate methods in quantum chemistry, but due to their extremely cost they can only be applied to small systems [14]. In this study, the CCSD(T) method, a truncated CC with first and second excitations and the third ones added perturbatively, has been used to compute single point energy calculations on the DFT optimized geometries for the systems in the absence of water ice and in the presence of the 3-water cluster model. For these calculations the Dunning aug-cc-pVTZ basis set has been used.

Truncated Coupled Cluster methods are the most accurate methods in quantum chemistry but due to its expensive cost they can only be applied to small systems. In this study CCSD(T) a truncated CC with first and second excitations and the third ones added perturbatively have been used to compute single point energy calculations of the optimized geometries with DFT functionals for the gas-phase scenario optimizations and the cluster model optimization. [14]

For the periodic calculations, additionally, the semi-empirical HF-3c method [15], which is based on the original HF combined with a MINIX basis set that include three corrections to alleviate the deficits caused by the approximations employed, was used to optimize the geometries. Subsequently, single point energy calculations on the HF-3c optimized geometries at the different DFT theory levels were carried out, the results of which were compared with those obtained at the full DFT level.

Interstellar water ice surfaces were modelled by adopting two different approaches: i) a cluster model made up by three water molecules, and ii) a crystalline periodic slab model. For this later case, the model is based on the crystalline P-ice structure. It has been generated by cutting out the P-ice 3D periodic bulk structure perpendicular to the [010] direction, resulting with the (010) slab surface model, which consists of twelve atomic layers (a thickness of 10.784 Å) [16], [17]. Additionally, different unit cells were considered aiming to study the effect of the lateral interactions between species belonging to adjacent cells: a 1x1cell, consisting of 24 water molecules (72 atoms), and a 2x2 cell, consisting of 96 water molecules (288 atoms).

## 3      Results

As mentioned above, the reactants, intermediate and products geometries have been optimized on the absence and in the presence of water ice.

In the absence of ice models (namely, a gas-phase reaction), the reactants, the intermediate and the products were optimized using the three different DFT functionals mentioned above: B3LYP-D3(BJ), PBE-D3(BJ) and M06-2X-D3. Additionally, CCSD(T) single point energy calculations on all the DFT optimized geometries were performed to check the accuracy of the DFT methods.

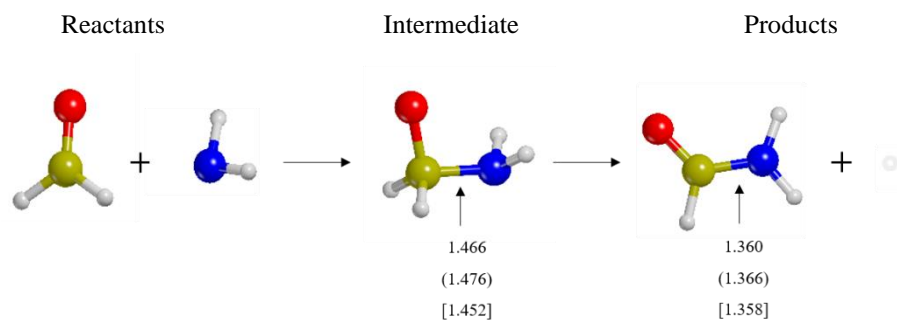

**Fig. 1.** Optimized structures of the reactants, intermediate and products of the studied reaction. Distances are in Å. Bare values correspond to those at B3LYP-D3(BJ), in brackets to those at PBE-D3(BJ) and in square brackets to those at M06-2X-D3. Atom color legend: red, O; blue, N; green, C white, H.

**Table 1.** Relative energies (in kJ/mol) of the stationary points with respect to the reactants for the reaction in absence of water ice computed at the different DFT functionals used for geometry optimizations and considering the single point energy calculation at CCSD(T).

| Geometry optimization functional | B3LYP-D3(BJ) | | PBE-D3(BJ) | | M06-2X-D3 | |
|---|---|---|---|---|---|---|
| Single point calculation | B3LYP-D3(BJ) | CCSD(T) | PBE-D3(BJ) | CCSD(T) | M06-2X-D3 | CCSD(T) |
| Reactants | 0.0 | 0.0 | 0.0 | 0.0 | 0.0 | 0.0 |
| Intermediate | -80.7 | -63.2 | -105.9 | -62.6 | -83.0 | -63.8 |
| Products | -43.4 | -37.2 | -77.4 | -37.1 | -59.0 | -37.3 |

Figure 1 shows the optimized stationary points at these DFT methods. Table 1 reports their relative energies with respect to the reactants, including the single point CCSD(T) values. Significant differences can be observed in the relative energies between the three DFT functionals, although changes in geometries are almost insignificant, less than 0.1 Å. Indeed, all DFT results indicate the intermediate as the most stable species along the path with some significant differences between DFT energy values specially with PBE-D3(BJ) results. In contrast, CCSD(T) energy values are almost in perfect agreement irrespective of the optimized geometry level and underline a general trend: the intermediate species is more stable than the products, in line with published works [5].

Aiming to characterize this formation route on water ice mantles, as a first step, geometry optimizations of the stationary points on a water trimer have been performed adopting the three DFT methods, which were subsequently computed at CCSD(T). Results are shown in Figure 2 and Table 2.

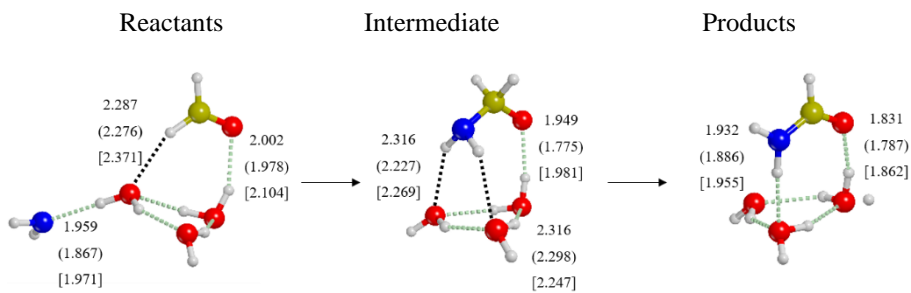

**Fig. 2**. Optimized structures of the reactants, intermediate and products of the studied reaction in the presence of a 3-$H_2O$ cluster. Distances are in Å. Bare values correspond to those at B3LYP-D3(BJ), in parenthesis to those at PBE-D3(BJ), and in brackets to those at M06-2X-D3.

**Table 2.** Relative energies (in kJ/mol) of the stationary points with respect to the reactants for the reaction under study in the presence of a 3-$H_2O$ cluster computed at the different DFT functionals and at CCSD(T) single point energy calculation on the DFT optimized geometries.

| Geometry optimization functional | B3LYP-D3(BJ) | | PBE-D3(BJ) | | M06-2X-D3 | |
|---|---|---|---|---|---|---|
| Single point calculation | B3LYP-D3(BJ) | CCSD(T) | PBE-D3(BJ) | CCSD(T) | M06-2X-D3 | CCSD(T) |
| Reactants | 0.0 | 0.0 | 0.0 | 0.0 | 0.0 | 0.0 |
| Intermediate | -64.1 | -53.5 | -92.6 | -32.8 | -74.4 | -55.0 |
| Products | -39.8 | -37.0 | -72.9 | -37.3 | -51.3 | -35.6 |

In this scenario, regarding the geometry optimizations, significant differences are found between optimizations at B3LYP-D3(BJ) and M06-2X-D3 compared with those at PBE-D3(BJ). Indeed, in this later case H-bond interactions connecting the $NH_2CH_2O$ species with the cluster are significantly shorter (by 0.2 Å) than those computed at the other levels (see Figure 2). Despite these geometry differences, relative energies indicate now that the intermediate and the products present a more similar stability than in the gas-phase scenario, the former species being somewhat more stable. However, single point CCSD(T) results do not follow the general trend observed in the absence of water ice, in which the intermediate was the most stable species of the route. In PBE-D3(BJ) optimized geometries case, the situation is reversed, giving the products more stable than the intermediate (see Table 2).

Characterization of the energetic features of this mechanistic path has also been performed on a more realistic ice surface by adopting a periodic approach. In this case, the geometry optimizations have been performed on the crystalline (010) slab surface model of P-ice using the three different DFT functionals and the cost-effective HF-3c method.

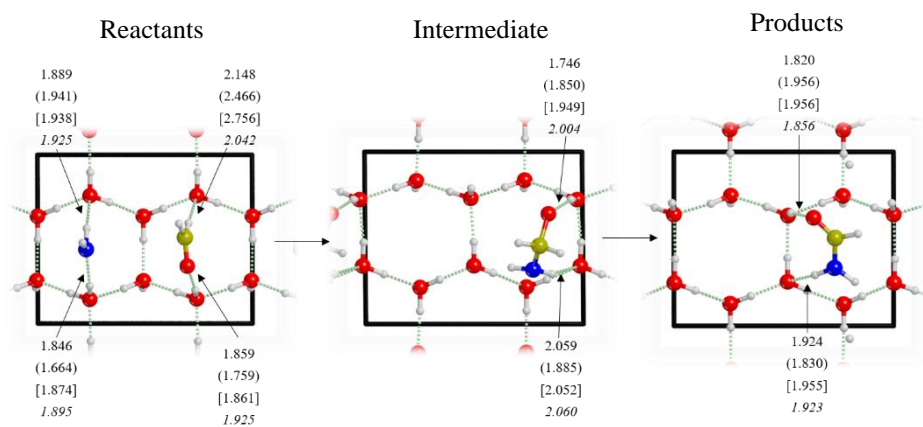

**Fig. 3**. Optimized structures of the reactants, intermediate and products of the studied reaction on the 1x1 crystalline water slab model. Distances are in Å. Bare values correspond to those at B3LYP-D3(BJ), in parenthesis to those at PBE-D3(BJ), in brackets to those at M06-2X-D3 and in italics to those at HF-3c.

**Table 3.** Relative energies (in kJ/mol) of the stationary points with respect to the reactants for the reaction under study on the 1x1 crystalline slab model computed at the different DFT functionals and at the HF-3c level and considering the single point energy calculation at the different DFT functionals on the optimized HF-3c geometries.

| Geometry optimization functional | B3LYP-D3(BJ) | PBE-D3(BJ) | M06-2X-D3 | HF-3c | | | |
|---|---|---|---|---|---|---|---|
| Single point calculation | B3LYP-D3(BJ) | PBE-D3(BJ) | M06-2X-D3 | HF-3c | B3LYP-D3(BJ) | PBE-D3(BJ) | M06-2X-D3 |
| Reactants | 0.0 | 0.0 | 0.0 | 0.0 | 0.0 | 0.0 | 0.0 |
| Intermediate | -19.1 | -42.4 | -26.1 | -113.6 | -15.7 | -34.6 | -31.8 |
| Products | -24.9 | -22.2 | -45.9 | 86.0 | -18.4 | -59.8 | -38.3 |

Optimized geometries of the stationary points on the 1x1 crystalline slab model are shown in Figure 3 and the computed energetics are reported in Table 3. Considering the energy values obtained with the three DFT functionals, it can be underlined that the B3LYP-D3(BJ) and the M06-2X-D3 functionals present the products as the most stable structures (see Table 3), predicting accordingly the reactions as exoergic processes. In contrast, this is not observed for the PBE-D3(BJ) functional, which is attributed to an overstabilization of the spin delocalization in the intermediate structures (usually in GGA functionals), as the spin density on the O atom of $NH_2CH_2O$ is 0.35 while for the other two functionals almost all the spin density is on the O atom.

Although HF-3c optimized geometries are in good agreement with DFT functionals, energy values are dramatically different, providing reaction energies largely endoergic. This is indicative that HF-3c is a very good cost/effective method for geometry optimizations but not for computing energetics. For this reason, single point energy calculations with the three DFT functionals have been carried out on the optimized HF-3c geometries. Obtained results clearly indicate that the reaction is exoergic, in which products is the most stable species, with no exception. Thus, these results indicate that the presence of water ice surfaces reverts the thermodynamics compared with the gas phase situation. This is probably due to the fact that the products ($NH_2CHO$ + H) are stabilised by favourable interactions with the surface (namely, H-bonds and dispersion), which are absent under strict gas-phase conditions. However, it is worth mentioning that the computed energetic values are significantly dependent on the functional method, in which the stability of the intermediate as well as the exoergicity of the processes follows the trend of (from more to less) PBE-D3(BJ) > M06-2X-D3 > B3LYP-D3(BJ). The significant stabilization of the intermediates and products exhibited by the PBE-D3(BJ) functional is probably due to its propensity to stabilize electron delocalized situations, as mentioned above. These results, additionally, indicate that, although energy values of HF-3c method are actually unreliable, performing DFT single point energy calculations on the HF-3c optimized geometries is a robust approach to obtain accurate results.

Finally, to assess possible lateral interaction effects between species belonging to adjacent cells (caused by the small unit cell size of the slab model), calculations on a slab model with a 2x2 unit cell (obtained by increasing by a factor 2 the *a* and *b* cell parameters) have been performed. In this case, to speed up the calculations, we optimized the systems at HF-3c followed by DFT single point calculations on them. Optimized geometries of the stationary points are shown in Figure 4 and the energetic values in Table 4.

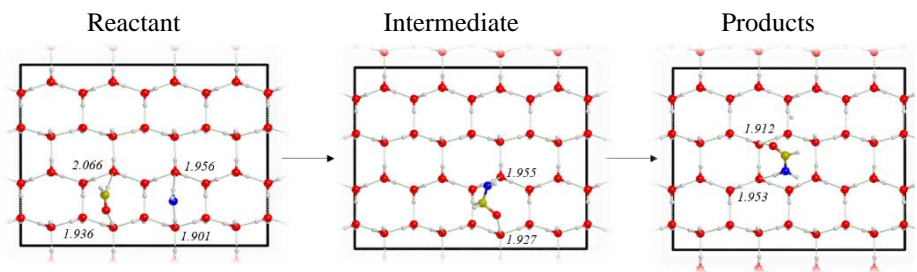

**Fig. 4**. HF-3c optimized structures of the reactants, intermediate and products of the studied reaction in on the 2x2 crystalline water slab model. Distances are in Å.

**Table 4.** Relative energies (in kJ/mol) of the stationary points with respect to the reactants for the studied reaction on the 2x2 crystalline slab model computed by considering single point energy calculations with the HF-3c method for geometry optimizations and considering DFT single point energy calculations on the HF-3c optimized geometries. Energetics calculated at HF-3c are also shown.

| Single point calculation | HF-3c | B3LYP-D3(BJ) | PBE-D3(BJ) | M06-2X-D3 |
|---|---|---|---|---|
| Reactants | 0.0 | 0.0 | 0.0 | 0.0 |
| Intermediate | -115.6 | -15.5 | -35.0 | -26.8 |
| Products | 85.4 | -19.0 | -50.1 | -36.3 |

As far as the geometries is concerned, no significant changes have been observed when they are compared (at HF-3c only) with those on the 1x1 crystalline slab model, variations being less than 0.1 Å. In the same way, computed energetics are also in line with those computed by adopting the 1x1 unit cell, all the cases indicating the products as the most stable species. Additionally, the trend of exoergicity of the reaction as a function of the DFT methods is kept. Accordingly, we can conclude that there are no lateral interaction effects and thus the 1x1 crystalline slab model is large enough to study the thermodynamics of this reaction by employing a periodic approach.

## 4 Conclusions

In this contribution, theoretical results relative to the formation of formamide by adopting the reaction of $NH_2 + H_2CO \rightarrow NH_2CHO + H$, which has been proposed to be a main synthetic route in the gas-phase, has been computed considering that it occurs on interstellar water ice surfaces. Water ice surfaces have been represented by modelled by adopting two different approaches: a water cluster consisting of three water molecules and a periodic surface of a crystalline water slab in which both 1x1 and 2x2 unit cell sizes have been used.

Simulations in absence of water ice and in the presence of the 3-$H_2O$ cluster model, have been performed using three different DFT functionals (B3LYP-D3(BJ), PBE-D3(BJ) and M06-2X-D3), in which, to check their accuracy, CCSD(T) single point energy calculations have been performed on the DFT optimized geometries. Simulations on the periodic 1x1 crystalline surface model, the same DFT methods have been performed, in which the semi-empirical HF-3c method has also been used to optimize the geometries followed by DFT single point calculations. On the periodic 2x2 crystalline model, this later methodology was only used to speed up the calculations.

In absence of water ice and in the presence of the 3-$H_2O$ cluster model the overall reaction is predicted to be endoergic, as the $NH_2CH_2O$ intermediate is more stable than the $NH_2CHO$ + H products. In contrast, on both the 1x1 and 2x2 periodic surfaces, for B3LYP-D3(BJ) and M06-2X-D3 results show that the final products are more stable than the intermediate species, indicating that the presence of the water surfaces revert the energetic trends observed in the gas phase and in the presence of the 3-$H_2O$ cluster. This is because the extended surfaces provide an extra-stability to the products due to the intermolecular forces (i.e., H-bond and dispersion) established with the "$NH_2CH_2O+$ H" species. Interestingly, this is not observed in PBE-D3(BJ) results because the functional, due to its definition as GGA functional, over stabilizes the intermediate species due to its propensity to favour electron delocalized situations.

Finally, the energetics provided by the HF-3c method are dramatically inaccurate. However, optimized geometries are similar to those at the DFT level and accordingly results from DFT single point energy calculations are very similar to those at full DFT level. Moreover, the unit cell size does not affect the results (i.e., those for the 1x1 and 2x2 unit cells are similar) and accordingly lateral interactions (if present) are not significant in the 1x1 unit cell, both surface models providing accurate thermodynamics for formamide formation on interstellar water icy grains.

This project has received funding from the European Research Council (ERC) under the European Union's Horizon 2020 research and innovation programme (grant agreement No. 865657) for the project "Quantum Chemistry on Interstellar Grains" (QUANTUMGRAIN). MINECO (project CTQ2017-89132-P) DIUE (project 2017SGR1323) are acknowledged for financial support. A.R. is indebted to the "Ramón y Cajal" program.